\documentstyle[12pt]{article}

\def\be{\begin{equation}}
\def\ee{\end{equation}}
\def\bea{\begin{eqnarray}}
\def\eea{\end{eqnarray}}
\def\SMM{Static Matrix Model}

\def\appendix#1{
  \addtocounter{section}{1}
  \setcounter{equation}{0}
  \renewcommand{\thesection}{\Alph{section}}
 \section*{Appendix \thesection\protect\indent \parbox[t]{11.715cm} {#1}}
  \addcontentsline{toc }{section}{Appendix \thesection\ \ \ #1}
  }

\def\e{{\, e}\,}

\def\be{\begin{equation}}
\def\ee{\end{equation}}
\def\beq{\begin{equation}}
\def\eeq{\end{equation}}
\def\bea{\begin{eqnarray}}
\def\eea{\end{eqnarray}}

\begin{document}

\setcounter{page}{0} 
\begin{flushright}
IPM-97-252 \\
hep-th/9711109
\end{flushright}

\pagestyle{plain} \vskip .25in

\begin{center}
{\Large {\bf M-Branes and Their Interactions in Static Matrix Model}\\ }
\end{center}
\vspace{.3cm}
\begin{center}
 
Amir H. Fatollahi$_{a,b}$\footnote{E-mail: fath@theory.ipm.ac.ir},
Kamran Kaviani$_{a,c}$\footnote{E-mail: kaviani@theory.ipm.ac.ir}, 
Shahrokh Parvizi$_{a}$\footnote{E-mail: parvizi@netware2.ipm.ac.ir} 

\vspace{.5 cm} 
{\it a)Institute for Studies in Theoretical Physics and Mathematics (IPM),}\\ 
{\it P.O.Box 19395-5531, Tehran, Iran.}\\ 

\vspace{.3 cm} 

{\it b)Department of Physics, Sharif University of Technology,\\ 
P.O.Box 11365-9161, Tehran, Iran.}\\ 
\vspace{.3 cm} 

{\it c)Department of Physics, Az-zahra University,\\
P.O.Box 19834, Tehran, Iran.}\\ 

\vskip .5 in 
{\bf ABSTRACT} 
\end{center}
Different BPS M-brane configurations including 
single and two parallel 
M$p$-branes ($p$= even) and M5-branes are introduced as the classical 
solutions of the recently proposed Static Matrix Model. 
Also the long range interactions of two relatively 
rotated M$p$-branes (one and two angles) and M$p$-brane--anti-M$p$-brane
are calculated. The results are in agreement with 11 dimensional 
supergravity results.


\newpage 
\setcounter{page}{1} \pagestyle{plain}

\section{Introduction}

M-theory \cite{T} played a central role in string theory unifications 
in the last two years. This approach to string theory unification 
become more concrete \cite{SS,M,BS,HW,BR} since the conjecture that
M-theory is a M(atrix) model \cite{BFSS}, a dimensional reduction of 
the 9+1 dimensional ${\cal N}=1$, $U(N)$ SYM to 0+1 dimension \cite{HALP}. 

In the recent paper \cite{AFKP} a Static Matrix Model was proposed 
for static configurations of M-theory. The main idea was based on 
the conjecture of \cite{BFSS} about equivalency of 
regularized supermembrane theory in the light-cone gauge and
M-theory in infinite momentum frame, which suggests generalization
of this equivalency to other gauges \cite{AFKP, FO}. 
The proposed model is defined in ten spatial dimensions 
and does not have any stringy parameter ($g_s,\;l_s,...$), but only 
the supermembrane tension $T_M$. Furthermore the model can not 
be considered as a dimensionally reduced SYM theory.

In \cite{AFKP} the long range interaction of M2-brane and anti-M2-brane
was calculated and found to be in agreement with the 11 dimensional 
supergravity results (i.e. $V(r)\sim {\frac{1}{{r^6}}}$). On the other hand, 
in \cite{AB} the result for the same 
problem appeared in compactified limit 
of the 11 dimensional supergravity (i.e. $V(r)\sim {\frac{1}{{r^5}}}$), 
in the context of the light cone M(atrix) model. This may be considered
as the advantage of {\SMM} which is defined in 10 spatial dimensions.

Here we would like to put the {\SMM} to further tests,
including interaction of rotated M$p$-branes, M$p$-branes and 
anti-M$p$-branes.

The article is organized as follows. Section 2 is devoted to a short 
review of {\SMM}. In section 3 we introduce some solutions of 
the equations of motion which preserve half of SUSY. In section 4 
we calculate long range interaction of M-branes including relatively 
rotated M-branes, with one or two angles, and M-anti-M-brane configurations.


\section{Static Matrix Model}

Here we give a short review of the \SMM \cite{AFKP}. 
The starting point is the supermembrane action in 11 dimensions 
\cite{DHN,BST}
\bea\label{1.1} 
S={\frac{-1}{{2}}} \int d^3\eta \bigg
(2\sqrt{-g} + \epsilon^{abc} \bar\theta \Gamma_{\mu\nu}\partial_a\theta
\times (\Pi_b^\mu p_c X^\nu +{\frac{1}{3}} \bar\theta\Gamma^\mu\partial_b%
\theta\; \bar\theta\Gamma^\nu\partial_c\theta)\bigg), \eea
\noindent where $\Pi$ and $g$ are 
\bea\label{1.5} 
\Pi_a^\mu&=& \partial_a
X^\mu + \bar\theta \Gamma^\mu \partial_a \theta, \nonumber\\ g_{ab}&=&\Pi_a
\cdot \Pi_b;
\eea
\noindent and $\theta$ is an eleven dimensional Majorana spinor (in this
section $a,b=0,1,2$).
We use the following notations everywhere:
$$
\mu ,\nu = 0,1,...,9,10;\;\;
I,J,K=1,2,...,9,10;       \;\;
i,j,k=1,2,...,9.
$$ 

By decomposition of the coordinates $\eta_a=(\tau, \sigma_r)$ , $r=1,2$,
and going to the static regime defined by 
\bea\label{1.10} 
X^0&\equiv&\tau, 
\nonumber\\ {\dot X}^I&\equiv& \dot\theta \equiv 0 ; 
\eea
\noindent the components of $g$ are found to be 
\bea\label{1.15}
g_{00}&=&-1, \nonumber\\ f_r\equiv
g_{0r}&=&\bar\theta\Gamma^0\partial_r\theta, \nonumber\\ g_{rs}&=&\bar{g}%
_{rs}-f_rf_s,\nonumber\\ \bar{g}_{rs}&\equiv& {\Pi_r}_I {\Pi_s}_I; 
\eea
\noindent and it can easily be shown that
\bea\label{1.20} g&=&-\bar{g},
\nonumber\\ \bar{g}&=&det\bar{g}_{rs}={\frac{1}{2}} \epsilon^{rs}%
\epsilon^{r's'} \bar{g}_{rr'}\bar{g}_{ss'} \nonumber\\ &=&{\frac{1}{2}}
(\epsilon^{rs} \Pi_r^I \Pi_s^J)^2. 
\eea

Putting all the above relations in (\ref{1.1}), we obtain 
\bea\label{1.25}
S= {\frac{1}{{2}}} \int d\tau\;d^2\sigma \bigg( -e^{-1} - e\;\bar{g} -2 {%
\epsilon}^{rs} \bar\theta\Gamma_{0I}\partial_r\theta \partial_s X^I \;- {%
\epsilon}^{rs} \bar\theta\Gamma_{0I}\partial_r\theta \;
\bar\theta\Gamma^I\partial_s\theta \bigg), 
\eea 
\noindent where $e$ appears as an auxiliary field for linearising the
action; its equation of motion gives 
\bea\label{1.30} 
e^2 \bar{g} \;=1, 
\eea
\noindent which can be used for eliminating $e$. Due to (\ref{1.30}),
configurations with $\bar{g}=0$ are unacceptable.

The action (\ref{1.1}) has a local fermionic $\kappa$-symmetry,
which allows one to gauge away half of the fermionic degrees of
freedom. $\theta$ is a 32-component 11-dimensional Majorana
spinor and in a real representation of $\Gamma$ matrices is real. We fix the 
$\kappa$-symmetry as for the light cone gauge, i.e. $
(\Gamma^0+\Gamma^{10})\theta=\Gamma^+\theta=0$. After integration over $
\tau$, the action (\ref{1.25}) takes the following form 
\be \label{1.35} 
S=-{\frac{1}{{2}}}{\cal {T}} \int d^2\sigma
e^{-1} \bigg(\;{\frac{1 }{2}} {\{X^i,X^j\}}^2+(\{X^i,X^{10}\}-{\frac{1}{2}}%
\lambda^T\{X^i,\lambda\})^2 +\lambda^T\gamma_i\{X^i,\lambda\}+1 \bigg), 
\ee 
\noindent where ${\cal {T}}$ is the time interval and 
$\lambda$ is the remaining 16 component part of $\theta$ and 
\be\label{1.40} 
\{ a,b \} =e\;(\partial_{{\sigma}_1}a\partial_{{\sigma}_2}b
-\partial_{{\sigma}_2}a\partial_{{\sigma}_1}b)= e\;\epsilon^{rs} \partial_ra
\partial_sb. \ee

By the known substitutions\cite{DHN,BFSS} 
\bea\label{1.45}
\{a,b\}&\Rightarrow& -i\; [ a,b ] ,\nonumber\\ \int \;e^{-1}\; d^2
\sigma&\Rightarrow& Tr, 
\eea 
\noindent one finds  
\bea\label{1.50} 
S=&-& \frac{1}{2}{\cal {T}}\; T_M^{5/3}\; Tr \;\bigg(\;{\frac{1 }{2}} \;
[X^i,X^j]^2+([X^i,X^{10}]-\frac{1}{2}\lambda^T[X^i,\lambda])^2
\;+\;i\lambda^T\gamma_i[X^i,\lambda]\bigg) \nonumber\\ 
&+& 6\pi {\cal {T}}\;T_M^{1/3}\; Tr \;(1), 
\eea 
\noindent in which $T_M$ is the supermembrane tension (in \cite{AFKP} the
numerical factors were fixed by interaction considerations).

The action (\ref{1.50}) has a gauge symmetry defined by: 
\bea\label{1.55}
\delta_{gauge}X^i&=&i[X^i,\alpha], \nonumber\\ 
\delta_{gauge}\lambda&=&i[%
\lambda,\alpha], \nonumber\\ 
\delta_{gauge}X^{10}&=&i[X^{10},\alpha]. 
\eea 

The action (\ref{1.50}) is invariant under the two SUSY 
transformations defined by two real anti-commutating $SO(9)$ spinors 
$\epsilon_1$ and $\epsilon_2$
\bea\label{1.60} 
\delta^{(1)}X^i&=& 0, \nonumber\\
\delta^{(1)}\lambda&=&\epsilon_1,     \nonumber\\
\delta^{(1)}(A^i A_i)&=&0\;\;\;(\Rightarrow 
\delta^{(1)}X^{10}={\frac{1}{{2}}}\epsilon_1^T\lambda),
\eea 
  and
\bea\label{65}
\delta^{(2)}X^i&=& \epsilon_2^T \gamma^I \lambda, \nonumber\\
\delta^{(2)}\lambda&=&\frac{i}{2} [X^i,X^j]\gamma_{ij}\epsilon_2,\nonumber\\
\delta^{(2)}(A^i A_i)&=&-i\lambda^T[\lambda,\lambda^T \epsilon_2]\;\;\;
(\Rightarrow \delta^{(2)}X^{10}=\cdot\cdot\cdot),
\eea
where 
$$
A^i=[X^i,X^{10}]-{\frac{1}{2}} \lambda^T[X^i,\lambda].
$$
The last line of these transformations can be used for defining
the SUSY transformations of $X^{10}$. The variations of $A^iA_i$
are the same as those of the auxiliary fields in SUSY theories.

Note that the balance between bosonic and fermionic degrees of freedom.
There are ten $X^I$; one of them, e.g. $X^{10}$ 
can be gauged away by the gauge symmetry (\ref{1.55}) and there remain 9 
bosonic degrees of freedom. 
Because of gauge fixing one complex ghost must be introduced 
; so there are 16+2=18(real)
fermionic degrees of freedom which gives the correct balance \cite{DHN}.


\section{BPS M-Brane Configurations}

\setcounter{equation}{0}
In this section we search for different BPS M-brane 
configurations of the model
which are described by solutions of the classical equations of motion. 
The classical equations of motion with  the condition $\lambda=0$ are 
\bea\label{2.1}
\sum_I [X^I,[X^I,X^J]]=0. 
\eea
\noindent Every configuration with $[X^I,X^J]\sim 1$ and 
with the other $X$'s vanishing are solutions of (\ref{2.1})
\footnote{
The point like (fully commuting) configurations which may be represented by 
$$ 
X^i = diag(x^i_1, x^i_2,...,  x^i_n),\;\;\;
X^{10}=\lambda=0,
$$
\noindent are not acceptable because of vanishing $\bar{g}$ in 
(\ref{1.30}). It is in agreement with the fact that the individual 
11 dimensional supergravitons which are candidates for "quark" 
substructure of our model (due to their role in infinite 
momentum frame  M(atrix) model as "partons") can not be studied as 
static configurations in 11 dimensions, because they are massless.

This argument also will be supported by the equation of motion of $n$, 
the size of matrices. By inserting solutions introduced above
in the action one finds,
$$
S=0\;+ n.
$$
The equation of motion for $n$ has no solutions (gives $1=0$).}.  

Solutions which can be interpreted as M$p$-brane have the form 
\be\label{2.15} 
X_I^{cl} = \left(B_1,B_2,\ldots,B_p,0,\ldots,0 \right), 
~~~~~\lambda=0\,, 
\ee
where $B_1,\ldots,B_p$ are $n\times n$ matrices with $n$ large, with
the commutation relations  
\be\label{2.20} 
[B_a,B_b]=ic_{ab}{\bf 1}\,, 
\ee
and $a,b=1,\ldots,p$, and for $2l\equiv p=2,4,6,8$
\footnote{
Interpretation of $p=4,6$ solutions as those found in \cite{G} 
is obscure, because of remaining SUSY in each case, these
preserve half of SUSY but those $\frac{1}{4}$ and $\frac{1}{6}$
for $p=4,6$ respectively. Although the "stack" behaviour of these solutions
suggests that the conditions introduced in \cite{G} on Killing spinors
may be reduced to one condition which preserves 
half of SUSY. By all of the above $p=8$ remains a problem \cite{BSS}.}.
            
By a proper rotation the anti--symmetric matrix $c_{ab}$ can be
brought to the Jordan form 
\begin{equation}\label{2.25} 
c_{ab}=\left( 
\begin{array}{ccccc}
0 & \omega_1 &  &  &  \\ 
-\omega_1 & 0 &  &  &  \\  
&  & \ddots &  &  \\  
&  &  & 0 & \omega_l \\  
&  &  & -\omega_l & 0 \\  
\end{array}
\right). 
\end{equation}

These solutions can be represented by ($2l=p$)
\begin{equation}
\label{2.35} \left\{ 
\begin{array}{lll}
X_{2i-1}&=&1_{n_1}\otimes 1_{n_2}\otimes ...\otimes 
\frac{L_{2i-1}}{\sqrt{2\pi n_i}}q_i
\otimes 1_{n_{i+1}}\otimes...\otimes 1_{n_l},\\ 
X_{2i}&=&1_{n_1}\otimes 1_{n_2}\otimes ...\otimes 
\frac{L_{2i}}{\sqrt{2\pi n_i}}p_i\otimes 1_{n_{i+1}}\otimes...
\otimes 1_{n_l}, \;\;\;\;\;\;l\geq i \\  
X^i &=&0,\;\;\;\;\;\;\;\;\;\; i > 2l=p,   
\end{array}
\right. 
\end{equation}
\noindent where $n_1 n_2...n_l=n$ and $L_a$'s are compactification radii,
with commutation relations 
$$
[ q_{i}, p_{j} ]= i\delta_{ij}\; 1_{n_i}.
$$
The eigenvalues of $q,\;p$ are uniformly distributed as 
$$
-\sqrt{\frac{\pi n_i}{2}}\leq q_i, p_i \leq \sqrt{\frac{\pi n_i}{2}}. 
$$
So the extension of solutions along $X_i$ axis is 
$L_i\rightarrow \infty$. 
Thus one can obtain 
\bea
[X_{2i-1}, X_{2i}]= {\frac{i}{2\pi n_i}} L_{2i-1} L_{2i} 1_n,
\;\;\;\;0\leq i\leq l,
\eea
and correspondingly
\begin{equation}\label{2.30} 
\frac{n^\frac{1}{l}}{L_{2i-1}L_{2i}} = \frac{1}{2\pi\omega_i},
\end{equation}
by $n_i\sim n^\frac{1}{l}$ \cite{BSS, FMOSZ}.

By putting these solutions in (\ref{1.50}) and assuming\cite{AFKP} 
\footnote{
The only relevant scale in the theory is 11 dimensional Planck length, 
$l_p, \; (T_M\sim l_p^{-3})$.}
$$
{\frac{L_{2i-1} L_{2i}}{2 \pi n_i}} \sim {T_M}^{\frac{-2}{3}}, 
$$
\noindent one finds 
\bea
S\sim {\cal {T}}\;T_M^{1/3} \;n \Rightarrow 
S&\sim &{\cal {T}}\;T_M^{1/3}L_1L_2...L_p T_M^{\frac{2l}{3}} , \nonumber\\ 
&\sim & T_p ({\cal {T}}L_1L_2...L_p), 
\eea
\noindent where the second line is the action of M$p$-brane after passing 
${\cal {T}}$ of time. So one finds 
$$
T_p\sim T_M^{\frac{{2l+1}}{3}}, 
$$
for the tension of a M$p$-brane.

Also one can introduce parallel M$p$-brane configurations. 
The configurations with two parallel M$p$-branes can be obtained from the
block-diagonal matrix with two identical blocks describing a pair of
M$p$-branes. Translating along the $(p+1)$-th axis by the distance $
r $ from each other we obtain the configuration of two parallel 
M$p$-branes ($2l\equiv p$)
\begin{eqnarray}\label{2.31}
 X_{a}^{cl}&=&\left(
 \begin{array}{cc}
 B_a & 0 \\
 0 & B_a \\
 \end{array}
 \right),~~~~a=1,\ldots,p,\nonumber \\ 
 X_{p+1}^{cl}&=&\left(
 \begin{array}{cc}
 \frac{r}{2} & 0 \\
 0 & -\frac{r}{2} \\
 \end{array}
 \right),\nonumber \\
 X_{10}^{cl}&=&X_{i }^{cl}=0,~~~~i =p+2,\ldots,9 \,.
 \end{eqnarray}

It can be shown that all of the above solutions 
preserve half of the SUSY: they are BPS.
By combining the two SUSY variations of $\lambda$ with 
the solutions as background(i.e. $[X_a,X_b]\sim c_{ab}$)
one can see that the solutions are invariant under half of the SUSY's
by requiring the condition
\bea\label{*}
\epsilon_1 \sim \frac{i}{2} c_{ab}\gamma_{ab}\epsilon_2
\eea
on the SUSY parameters $\epsilon_{1,2}$. The rank of the 
$16\times 16$ matrix which appeared in RHS of (\ref{*}) is 8 and 4 
in cases $p=4,6$ respectively. 
This supports the argument of the previous footnote  
about identification of these solutions and those of \cite{G}. 

The above argument is in agreement with the one-loop effective action 
considerations. The one-loop effective action is calculated with 
the above backgrounds in \cite{AFKP} 
\be\label{2.40} 
\;W= \;{\frac{1}{2}}Trlog\bigg(
P_K^2\delta_{IJ}-2iF_{IJ}\bigg)- {\frac{1 }{4}}\;
Trlog\bigg((P_I^2+{\frac{i}{{2}}}\; F_{IJ}\Gamma^{IJ})
(\frac{1+\Gamma_{11}}{2})\bigg)-Trlog(P_I^2), \nonumber\\
\ee       
\noindent with $P_I\;*=[X_I^{cl},*]$, $F_{IJ}\;*=[f_{IJ},*]$, $
f_{IJ}=i[X_I^{cl},X_J^{cl}]$.

For the above solutions one has $F_{IJ}=0$, so
\bea\label{2.45}
W\sim \;(\frac{1}{2}\cdot 10- \frac{1}{4}\cdot 16-1)\;Trlog(P_I^2)=0.
\eea
This indicates that solutions won't take quantum corrections, as one
expects for a BPS state.

Before ending this section it is interesting to consider possible odd  
dimensional solutions. 
For example in 5 dimensions, a solution can be represented by:
\begin{equation}
\label{2.115} \left\{ 
\begin{array}{lll}
X_1= \frac{L_1}{\sqrt{2\pi n_1}}q_1\otimes 1_{n_{2}}\otimes 1_{n_{3}},\\ 
X_{2}=\frac{L_2}{\sqrt{2\pi n_1}}p_1\otimes 1_{n_{2}}\otimes 1_{n_{3}}, \\  
X_{3}=1_{n_{1}}\otimes \frac{L_3}{\sqrt{2\pi n_2}}q_2\otimes 1_{n_{3}},\\
X_4=1_{n_{1}}\otimes \frac{L_4}{\sqrt{2\pi n_2}}p_2\otimes 1_{n_{3}},\\
X_5=1_{n_{1}}\otimes 1_{n_{2}}\otimes \frac{L_5}{(2\pi n_3)^x} q_3,\\
X_i=X_{10}=0,\;\;\;\;\;\;i>5,
\end{array}
\right. 
\end{equation}
where $n_1 n_2 n_3=n$, and we have let the power $x$ in $X_5$ be
unknown to be determined later. 
The desired commutation relations are
$$
[ q_{i}, p_{i} ]= i 1_{n_i};\;\;\; 
[ q_{3}, q_{i} ]=[ q_{3}, p_{i} ]=0,\;\;\;i=1,2,
$$
and the following eigenvalue distributions
$$
-\sqrt{\frac{\pi n_i}{2}}\leq q_i, p_i \leq \sqrt{\frac{\pi n_i}{2}}, 
$$
$$
-({\frac{\pi n_3}{2}})^x\leq q_3 \leq ({\frac{\pi n_3}{2}})^x. 
$$
So the extensions of the solution in 12345 directions are $L_{12345}$.
To understanding this solution further it is convenient 
to diagonalize $q_3$. The solution can be 
interpreted by two infinite stacks of M2-branes as a 4 dimensional object, 
stacked along 5th direction. By ignoring the 5th direction one notes the
similar stack behaviour of this solution and the longitudinal 
5-branes of M(atrix) theory \cite{BSS}
\footnote{
Solutions like (\ref{2.115}) can not be interpreted as 5 
dimensional objects in the context of the light cone M(atrix) theory. 
This can be understood
by studying the related supercharge which was found in \cite{BSS}, 
$$
Z_{1234}\sim R_{11} L_1L_2L_3L_4,
$$
which indicates extension along the longitudinal direction, $R_{11}$. 
So by considering the 5th direction one finds a 6 dimensional object.
}.
By putting this solution in (\ref{1.50}) one finds:
\bea
S\sim {\cal {T}}\;T_M^{1/3}\;n 
\Rightarrow S&\sim &{\cal {T}}\;T_M^{1/3}L_1L_2L_3L_4 L_5^\frac{1}{x} \;
T_M^{\frac{4x+1}{3x}}, \nonumber\\ 
&\sim & T_5 ({\cal {T}}L_1L_2L_3L_4 L_5), 
\eea
\noindent where the second line is the action of M5-brane extended 
in the 12345 directions and after passing ${\cal {T}}$ of time. 
Therfore, tension becomes
$$
T_5\sim T_M^\frac{5x+1}{3x}\; L_5^\frac{1-x}{x}.
$$
To have finite and non-zero tension as $L_5\rightarrow 
\infty$ one must put $x=1$. Then the tension is found to 
be 
$$
T_5\sim T_M^2,
$$
as one expected for a M5-brane \cite{SCH}. 

The other simple solution, which is consistent with the 
equation of motion of $n$, is a 3 dimensional solution represented by
\begin{equation}\label{2.100} 
\left\{ 
\begin{array}{lll}
X_1= \frac{L_1}{\sqrt{2\pi n_1}}q_1\otimes 1_{n_{2}},\\ 
X_{2}=\frac{L_2}{\sqrt{2\pi n_1}}p_1\otimes 1_{n_{2}}, \\  
X_{3}=1_{n_{1}}\otimes \frac{L_3}{2\pi n_2}q_2,\\
X^i =X_{10}=0, \;\;\;\;\;\;\;\;\;\; i > 3,   
\end{array}
\right. 
\end{equation}
where $n_1 n_2=n$ and 
$$
[ q_{1}, p_{1} ]= i 1_{n_1};\;\;\;
[ q_{2}, q_{1} ]=[ q_{2}, p_{1} ]=0.
$$
The operators $q_1$ and $p_1$ represent 
extensions in two directions and continuous spectrum of $q_2$ represents 
an infinite continuous "stack" of two dimensional objects along
the 3rd direction.


\section{M-Brane Long Range Interactions}

\setcounter{equation}{0}
In this section we calculate the long range interaction between different
M-brane configurations
\footnote{
Because of some similarities, at one loop order, 
between {\SMM} and the matrix theory approach 
to the type IIB superstring theory (called IKKT) \cite{IKKT}, there is 
a short cut for us to use the techniques developed in that framework
\cite{IKKT,FMOSZ,CMZFS}. Although for completion we try to give the 
required details.}.

The one-loop effective action $W$ was introduced in the previous section
(and calculated in \cite{AFKP}) with the backgrounds 
$\lambda=X_{10}^{cl}=0$, (writing in ten dimensional language)
\be \label{3.1} 
\;W= \;{\frac{1}{2}}Trlog\bigg(
P_K^2\delta_{IJ}-2iF_{IJ}\bigg)- {\frac{1 }{4}}\;
Trlog\bigg((P_I^2+{\frac{i}{{2}}}\; F_{IJ}\Gamma^{IJ})
(\frac{1+\Gamma_{11}}{2})\bigg)-Trlog(P_I^2), \nonumber\\
\ee       
with $P_I\;*=[X_I^{cl},*]$, $F_{IJ}\;*=[f_{IJ},*]$, 
$f_{IJ}=i[X_I^{cl},X_J^{cl}]$ and 
$$
\Gamma^i_{32}= \left( 
\matrix{0 & \gamma^i_{16}  \cr
\gamma^i_{16} & 0 }\right),
\;\;\;
\Gamma^{10}_{32}= \left( 
\matrix{1_{16} & 0  \cr
0 & -1_{16} }\right),\;\;\Gamma_{11}=i\Gamma_{123456789,10}.
$$

In the cases of our interest we have $[X_I^{cl}, f_{IJ}]=c-number$, and 
so $P_I^2$ and $F_{IJ}$ are simultaneously diagonalizable.
By setting $F_{IJ}$ to be in the Jordan form 
(because of $X_{10}^{cl}=0$ we put $a_5=0$)

\begin{equation}\label{3.5} 
F_{IJ}=\left( 
\begin{array}{ccccccc}
0  &-a_1&      &   &    &       &          \\ 
a_1&  0 &      &   &    &       &          \\  
   &    &\ddots&   &    &       &          \\  
   &    &      & 0 &-a_4&       &          \\  
   &    &      &a_4&  0 &       &          \\  
   &    &      &   &    &    0  & -a_5(=0) \\
   &    &      &   &    &a_5(=0)&   0      \\
\end{array}
\right), 
\end{equation}
and noting that eigenvalues of $\Gamma^{IJ}$'s are $\pm i$ 
one finds the following expression for $W$ \cite{IKKT}
\bea\label{3.10}
W=&\frac{1}{2}&\sum_{i=1}^{5}Trlog(1-\frac{4a_i^2}{(P_I^2)^2})\nonumber\\
  &-\frac{1}{4}&
\sum_{
 \begin{array}{c}
 \scriptstyle
 s_1,\ldots,s_5=\pm 1\\[-2.5mm]
 \scriptstyle
 s_1\ldots s_5=1
 \end{array}}
  Trlog(1-\frac{a_1s_1+...+a_5s_5}{P_I^2}),
\eea
where the sum over $s_a$'s is restricted because of projecting 
on 16 dimensional  subspace of $\Gamma$ matrices by tracing only on that 
part of basis which give 1 by acting the operator
$i\Gamma^{12}\Gamma^{34}\Gamma^{56}\Gamma^{78}\Gamma^{9,10}$.


\subsection{Rotated M-Branes Interaction (two angles)}

The configuration with two rotated M$p$-branes can be obtained from the
block-diagonal matrix with two identical blocks describing a pair of
M$p$-branes. Translating along the $(p+3)$-th axis by the distance $
r $ from each other and rotating in opposite directions in $(p,p+1)$ plane
and $(p-1,p+2)$ through the angles $\phi /2$ and $\psi /2$, we obtain the
configuration of two rotated M$p$-branes ($2l\equiv p$=2,4,6)
\begin{eqnarray}\label{3.50}
 X_{a}^{cl}&=&\left(
 \begin{array}{cc}
 B_a & 0 \\
 0 & B_a \\
 \end{array}
 \right),~~~~a=1,\ldots,p-2,\nonumber \\ 
 X_{p-1}^{cl}&=&\left( 
 \begin{array}{cc}
 B_{p-1}\cos\frac{\psi }{2} & 0 \\
 0 & B_{p-1}\cos\frac{\psi }{2} \\
 \end{array}
 \right),    \nonumber \\
 X_{p}^{cl}&=&\left(
 \begin{array}{cc}
 B_p\cos\frac{\phi }{2} & 0 \\
 0 & B_p\cos\frac{\phi }{2} \\
 \end{array}
 \right),      \nonumber \\ 
 X_{p+1}^{cl}&=&\left(
 \begin{array}{cc}
 B_p\sin\frac{\phi }{2} & 0 \\
 0 & -B_p\sin\frac{\phi }{2} \\
 \end{array}
 \right),          \nonumber \\ 
 X_{p+2}^{cl}&=&\left(
 \begin{array}{cc}
 B_{p-1}\sin\frac{\psi }{2} & 0 \\
 0 & -B_{p-1}\sin\frac{\psi }{2} \\
 \end{array}
 \right),          \nonumber \\
 X_{p+3}^{cl}&=&\left(
 \begin{array}{cc}
 \frac{r}{2} & 0 \\
 0 & -\frac{r}{2} \\
 \end{array}
 \right),\nonumber \\
 X_{10}^{cl}&=&X_{i }^{cl}=0,~~~~i =p+4,\ldots,9 \,.
 \end{eqnarray}

So one finds 
\bea\label{3.55}
f_{p-1,p}&=&-\omega_l\cos\frac{\psi}{2}\;\cos\frac{\phi}{2}\otimes 1,
\;\;\;\;\;
f_{p-1,p+1}=-\omega_l\cos\frac{\psi}{2}\;\sin\frac{\phi}{2}\otimes\sigma_3,
\nonumber\\ 
f_{p,p+2}&=&\omega_l\cos\frac{\phi}{2}\;\sin\frac{\psi}{2}\otimes\sigma_3,
\;\;\;\;\;
f_{p+1,p+2}=\omega_l\sin\frac{\phi}{2}\;\sin\frac{\psi}{2}\otimes 1,
\nonumber\\
{\rm otherwise}\;\;\;f_{ab}&=&0,
\eea
where $\sigma_3$ is the Pauli matrix. Then
\bea\label{3.60}
[P_{p-1},P_{p+1}]&=&i\omega_l\cos\frac{\psi}{2}\;\sin\frac{\phi}{2}
\otimes\Sigma_3,\nonumber\\
{}[P_{p},P_{p+2}]&=&-i\omega_l\cos\frac{\phi}{2}\;\sin\frac{\psi}{2}
\otimes\Sigma_3,\nonumber\\
{}P_{p+3}&=&\frac{r}{2}\otimes\Sigma_3,\nonumber\\   
{}{\rm otherwise}\;\;\;F_{ab}&=&0,   
\eea
where 
$$
\Sigma_3 *=[1\otimes\sigma_3,*].
$$
$\Sigma_3$ has 2, -2, 0, 0 as eigenvalues. Zero eigenvalues won't have any
contribution to the effective action because of (\ref{2.45}). The 
other two force $P_{p-1},P_{p+1}$ and $P_{p},P_{p+2}$ to behave as harmonic 
oscillators with their related frequencies to be read from (\ref{3.60}).
So the eigenvalues of $P_I^2$ are 
\bea\label{3.65} 
E_{{\bf q},{\bf p},k,k^{\prime}}
=r^2+2\sum_{i=1}^{l-1}(q_i^2+p_i^2) 
+4\omega_l\cos\frac{\psi}{2}\;\sin\frac{\phi}{2}(k+\frac{1}{2})+
4\omega_l\sin\frac{\psi}{2}\;\cos\frac{\phi}{2}(k'+\frac{1}{2}),\nonumber\\
\eea
where $q_i,p_i$ are eigenvalues of $P_i$; $a=1,..,p-2$ and $k,k'$ are
harmonic oscillator numbers.

So one finds for (\ref{3.10})
\begin{eqnarray}\label{3.70}
W=&~&\prod_{a=1}^{p-2}
 \left(
 \frac{n^{\frac{1}{l}}}{L_a}
 \right)
 \int d^{l-1}q\,d^{l-1}p  \times
 \nonumber\\
&~&\sum_{k',k=0}^{\infty }\left[
 \ln\left(1-\frac{16\omega^2_l\cos^2{\frac{\psi}{2}}\sin^2{\phi\over 2}}
 {E_{{\bf q},{\bf p},k,k'}^2}\right)
 +\ln\left(1-\frac{16\omega^2_l\cos^2{\frac{\phi}{2}}\sin^2{\psi\over 2}}
 {E_{{\bf q},{\bf p},k,k'}^2}\right)
\vphantom{-\frac{1}{2}
\sum_{
 \begin{array}{c}
 \scriptstyle
 s_1,\ldots,s_5=\pm 1\\[-2.5mm]
 \scriptstyle
 s_1\ldots s_5=1
 \end{array}}
 \ln\left(1-\frac{2\omega _1s_1}{E_{{\bf q},{\bf p},k,k'}}
 \right)}
 \right.\nonumber\\
&~&\left.
 -\,\frac{1}{2}\sum_{
 \begin{array}{c}
 \scriptstyle
 s_1,\ldots,s_5=\pm 1\\[-2.5mm]
 \scriptstyle
 s_1\ldots s_5=1
 \end{array}}
 \ln\left(1-\frac{2s_1\omega_l\cos{\psi\over 2}\sin{\phi\over 2}+
 2s_2\omega_l\cos{\phi\over 2}\sin{\psi\over 2} }
 {E_{{\bf q},{\bf p},k,k'}}
 \right)
 \right].
 \end{eqnarray}
The last term in the above can be rewritten as 
\bea
&{}&
\vphantom{\frac{1}{2}\sum_{
 \begin{array}{c}
 \scriptstyle
 s_1,\ldots,s_5=\pm 1\\[-2.5mm]
 \scriptstyle
 s_1\ldots s_5=1
 \end{array}
 }\ln\left(1-\frac{2\omega _1s_1}{E_{{\bf q},{\bf p},k,k'}}
 \right)} \,\frac{1}{2}\sum_{
\begin{array}{c}
\scriptstyle  s_1,\ldots,s_5=\pm 1 \\ [-2.5mm] \scriptstyle  s_1\ldots s_5=1 
\end{array}}
\ln\left(1-\frac{2s_1\omega_l\cos{\psi\over 2}\sin{\phi\over 2}+
2s_2\omega_l\cos{\phi\over 2}\sin{\psi\over 2}} {E_{{\bf q},{\bf p}
,k,k'}} \right) = 
\nonumber\\ 
&{}&2 \ln\left(1-\frac{4\omega_l^2\sin^2{\frac{\phi+\psi}{2}}}
{E^2_{{\bf q},{\bf p},k,k'}} \right)+ 
2 \ln\left(1-\frac{4\omega_l^2\sin^2{\frac{\phi-\psi}{2}}}
{E^2_{{\bf q},{\bf p},k,k'}} \right).
\eea

It is convenient to represent the logarithms in (\ref{3.70}) in the form: 
\begin{equation}\label{3.75} 
\ln \frac uv =\int\limits_0^{\infty}\frac{ds}{s}\left(\e^{-vs}
- \e^{-us} \right). 
\end{equation}
The sums over $k,\;k^{\prime}$ and the integrals over $q$ and $p$  can
be calculated as follows 
\bea\label{3.80} 
&~&\int d^{l-1}q\,d^{l-1}p\, \sum_{k,k^{\prime}=0}^{\infty }\e^{-sE_{%
{\bf q},{\bf p},k,k^{\prime}}} =\nonumber\\
&~&\left(\frac{\pi }{2s}\right)^{\frac{p}{2}-1} 
\frac{\e^{-r^2s}}
{
4\sinh\left({2s\omega_l\cos\frac{\psi}{2} \sin\frac{\phi}{2}}\right)\; 
\sinh\left({2s\omega_l\sin\frac{\psi}{2}\cos\frac{\phi}{2}}\right)}
\eea

We finally obtain the following form 
\bea\label{3.85} 
W=\prod_{a=1}^{p-2}
 \left(
 \frac{n^{\frac{1}{l}}}{L_a}
 \right)
\int_0^{\infty} 
\frac{ds}{s}\left ( \frac{\pi}{2s} \right )^{\frac{p}{2}-1} e^{-r^2s} \times
\nonumber\\
\left[
\frac
{2\sinh^2(\omega_ls\sin\frac{\phi+\psi}{2}) 
+2\sinh^2(\omega_ls\sin\frac{\phi-\psi}{2})}
{\sinh(2\omega_ls\sin\frac{\phi}{2}\cos\frac{\psi}{2}) 
\sinh(2\omega_ls\sin\frac{\psi}{2}\cos\frac{\phi}{2})}\;-\right.
\nonumber\\
\left.
\frac
{\sinh^2(2\omega_ls\sin\frac{\phi}{2}\cos\frac{\psi}{2}) 
+\sinh^2(2\omega_ls\sin\frac{\psi}{2}\cos\frac{\phi}{2})}
{\sinh(2\omega_ls\sin\frac{\phi}{2}\cos\frac{\psi}{2}) 
\sinh(2\omega_ls\sin\frac{\psi}{2}\cos\frac{\phi}{2})}
\right].
\eea

For large separation between branes we find: 
\bea\label{3.100} 
W&=&\frac{2}{3}(\frac{\pi}{2})^{l-1}\omega_l^2(2-l)!
\prod_{a=1}^{p-2}
 \left(
 \frac{n^{\frac{1}{l}}}{L_a}
 \right)\times
\nonumber\\
&~&\frac{
\sin^4\frac{\phi+\psi}{2}+\sin^4\frac{\phi-\psi}{2}
-8\cos^4\frac{\psi}{2}\sin^4\frac{\phi}{2}
-8\cos^4\frac{\phi}{2}\sin^4\frac{\psi}{2}  }{
\sin\psi\sin\phi}
\left(\frac{1}{r^{6-p}}\right)+\cdot\cdot\cdot   \nonumber\\
&=&-(\frac{\pi}{2})^{l-1}\omega_l^2(2-l)!
\prod_{a=1}^{p-2}
 \left(
 \frac{n^{\frac{1}{l}}}{L_a}
 \right)\times
\frac{(\cos\phi-\cos\psi)^2}{\sin\psi\sin\phi} 
\left(\frac{1}{r^{6-p}}\right)+\cdot\cdot\cdot ,
\eea
which is in agreement with supergravity results both in angular and 
$r$  dependences \cite{SJ}.

The above interaction vanishes in $\psi=\phi$ cases, signalling 
enhancement of SUSY. An equivalent result is obtained in
\cite{SJ} by considering interactions of rotated D$p$-branes, 
and in \cite{TI} by studying the SUSY algebra for rotated M-objects.


\subsection{Rotated M-Branes Interaction (one angle)}

By putting $\psi=0$ in two rotated angle configurations of the previous 
subsection one finds the following for one angle case:  
\begin{eqnarray}\label{3.150}
 X_{a}^{cl}&=&\left(
 \begin{array}{cc}
 B_a & 0 \\
 0 & B_a \\
 \end{array}
 \right),~~~~a=1,\ldots,p-1,\nonumber \\ 
 X_{p}^{cl}&=&\left(
 \begin{array}{cc}
 B_p\cos\frac{\phi }{2} & 0 \\
 0 & B_p\cos\frac{\phi }{2} \\
 \end{array}
 \right),      \nonumber \\ 
 X_{p+1}^{cl}&=&\left(
 \begin{array}{cc}
 B_p\sin\frac{\phi }{2} & 0 \\
 0 & -B_p\sin\frac{\phi }{2} \\
 \end{array}
 \right),          \nonumber \\ 
 X_{p+2}^{cl}&=&\left(
 \begin{array}{cc}
 \frac{r}{2} & 0 \\
 0 & -\frac{r}{2} \\
 \end{array}
 \right),\nonumber \\
 X_{10}^{cl}&=&X_{i }^{cl}=0,~~~~i =p+3,\ldots,9 \,.
 \end{eqnarray}

So one finds
\bea\label{3.155}
f_{p-1,p}&=&-\omega_l\cos\frac{\phi}{2}\otimes 1,\nonumber\\
f_{p-1,p+1}&=&-\omega_l\sin\frac{\phi}{2}\otimes\sigma_3,\nonumber\\ 
{\rm otherwise}\;\;\;f_{ab}&=&0,
\eea
and then
\bea\label{3.160}
[P_{p-1},P_{p+1}]&=&i\omega_l\sin\frac{\phi}{2}\otimes\Sigma_3,\nonumber\\
{}P_{p+3}&=&\frac{r}{2}\otimes\Sigma_3,\nonumber\\   
{\rm otherwise} \;\;\;{}F_{ab}&=&0.
\eea
Again we have an harmonic oscillator and the eigenvalues of $P^2_I$ are 
\begin{equation}\label{3.165} 
E_{{\bf q},{\bf p},k}=r^2+2\sum_{i=1}^{l-1}(q_i^2+p_i^2)+2
\cos^2 \frac{\phi}{2}\,~q_l^2 +4\omega_l\sin\frac{\phi }{2}\,~
(k+\frac{1}{2})\,. 
\end{equation}

So one finds for (\ref{3.10}) 
\begin{eqnarray}\label{3.170}
 W&=&
 \prod_{a=1}^{p-1}
 \left(
 \frac{n^{\frac{1}{l}}}{L_a}
 \right)
 \int d^lq\,d^{l-1}p\,
 \sum_{k=0}^{\infty }\left[
 \ln\left(1-\frac{16\omega^2 _l\sin^2 {\phi\over 2}}
 {E_{{\bf q},{\bf p},k}^2}\right)
 \vphantom{-\frac{1}{2}\sum_{
 \begin{array}{c}
 \scriptstyle
 s_1,\ldots,s_5=\pm 1\\[-2.5mm]
 \scriptstyle
 s_1\ldots s_5=1
 \end{array}
 }\ln\left(1-\frac{2\omega _1s_l}{E_{{\bf q},{\bf p},k}}
 \right)}
 \right.\nonumber \\&&\left.
 -\,\frac{1}{2}\sum_{
 \begin{array}{c}
 \scriptstyle
 s_1,\ldots,s_5=\pm 1\\[-2.5mm]
 \scriptstyle
 s_1\ldots s_5=1
 \end{array}
 }\ln\left(1-\frac{2s_1\omega _l\sin{\phi\over 2}}
 {E_{{\bf q},{\bf p},k}}
 \right)
 \right].
 \end{eqnarray}
The last term which originates from the integration over fermions can be
rewritten as \be
\vphantom{\frac{1}{2}\sum_{
 \begin{array}{c}
 \scriptstyle
 s_1,\ldots,s_5=\pm 1\\[-2.5mm]
 \scriptstyle
 s_1\ldots s_5=1
 \end{array}
 }\ln\left(1-\frac{2\omega_ls_1}{E_{{\bf q},{\bf p},k}}
 \right)} \,\frac{1}{2}\sum_{
\begin{array}{c}
\scriptstyle  s_1,\ldots,s_5=\pm 1 \\ [-2.5mm] \scriptstyle  s_1\ldots s_5=1 
\end{array}
}\ln\left(1-\frac{2s_1\omega_l\sin{\frac{\phi}{2}}} {E_{{\bf q},{\bf p},k}}
\right) = 4 \ln\left(1-\frac{4\omega_l^2\sin^2{\frac{\phi}{2}}} 
{E^2_{{\bf q},{\bf p},k}} \right). 
\ee

The sum over $k$ and the integrals over $q$ and $p$ then can be evaluated
using the formula 
\begin{equation}\label{3.200} 
\int d^lq\,d^{l-1}p\, \sum_{k=0}^{\infty }\e^{-sE_{{\bf q},{\bf 
p},k}} =\left(\frac{\pi }{2s}\right)^{\frac{p-1}{2}}\frac{\e^{-r^2s}}
{2\cos{\frac{\phi}{2}}\, \sinh\left(2\omega_ls\sin{\frac{\phi}{2}}\right)}. 
\end{equation}

We finally obtain the following form 
\begin{equation}\label{3.205} 
W= 
 \prod_{a=1}^{p-1}
 \left(
 \frac{n^{\frac{1}{l}}}{L_a}
 \right)
\int_0^{\infty} 
\frac{ds}{s}\left ( \frac{\pi}{2s} \right )^{\frac{p-1}{2}} e^{-r^2s} 
\frac
{4\sinh^2(\omega_ls\sin{\frac{\phi}{2}}) - 
\sinh^2(2\omega_ls\sin{\frac{\phi}{2}})} 
{\cos{\frac{\phi}{2}}~\sinh(2\omega_ls\sin{\frac{\phi}{2}})}\,. 
\end{equation}

For large separation between branes we find: 
\begin{equation}\label{220}
W=-4 (\frac{\pi}{2})^{\frac{p-1}{2}}  \,(5/2-l)!\,
\prod_{a=1}^{p-1}
 \left(
 \frac{n^{\frac{1}{l}}}{L_a}
 \right)
\omega_l^3
\tan\frac{\phi}{2}\,\sin^2\frac{\phi}{2}\,\frac 1{r^{7-p}}+
\cdot\cdot\cdot ,
\end{equation}
again in agreement with supergravity results both in angular and 
in $r$ dependences \cite{SJ}.

\subsection{M-Brane and Anti-M-Brane Interaction}

The M2-brane and anti-M2-brane long range interaction 
have been studied in the framework of M(atrix) theory in 
\cite{AB, LM} which their results appeared in compactified limit of
11 dimensional supergravity (i.e. $V(r)\sim \frac{1}{r^{5}}$).
Here we want to consider the same problem but for
M$p$-brane and anti-M$p$-brane in the context of our model. 
The long range interaction for two anti-parallel M$p$-branes was 
calculated in \cite{AFKP} for M2-branes.

The classical solution describing anti-parallel M$p$-branes at 
the distance $r$ from each other are represented by block--diagonal 
matrices (this solution can be obtained by setting $\phi=\pi$ 
in the previous subsection)
\begin{eqnarray}\label{3.260}
 X_{a}^{cl}&=&\left(
 \begin{array}{cc}
 B_a & 0 \\
 0 & B_a \\
 \end{array}
 \right),~~~~a=1,\ldots,p-1
 \nonumber \\
 X_{p}^{cl}&=&\left(
 \begin{array}{cc}
 B_a & 0 \\
 0 & -B_a \\
 \end{array}
 \right),
 \nonumber\\
 X_{p+1}^{cl}&=&\left(
 \begin{array}{cc}
 r/2 & 0 \\
 0 & -r/2 \\
 \end{array}
 \right),\nonumber \\
 X_{i }^{cl}&=&0,~~~~i =p+2,\ldots,9,
 \end{eqnarray}

So one finds
\bea\label{3.265}
f_{p-1,p}&=&-\omega_l\otimes \sigma_3,\;\;
\nonumber\\ 
{\rm otherwise}\;\;\;f_{ab}&=&0,
\eea
and then
\bea\label{3.270}
[P_{p-1},P_{p}]&=&i\omega_l\otimes\Sigma_3,\nonumber\\
{}P_{p+3}&=&\frac{r}{2}\otimes\Sigma_3,\nonumber\\   
{\rm otherwise} \;\;\;{}F_{ab}&=&0.
\eea
Again we have harmonic oscillators and the eigenvalues of $P^2_I$ are 
\begin{equation}\label{3.275} 
E_{{\bf q},{\bf p},k}=r^2+2\sum_{i=1}^{l-1}(q_i^2+p_i^2)+
4\omega_l\;
(k+\frac{1}{2})\,. 
\end{equation}

So one finds for (\ref{3.10}) 
\begin{eqnarray}\label{3.280}
 W&=&
 \prod_{a=1}^{p-1}\left(
 \frac{n^{\frac{1}{l}}}{L_a}
 \right)
 \int d^{l-1}q\,d^{l-1}p\,
 \sum_{k=0}^{\infty }\left[
 \ln\left(1-\frac{16\omega^2_l}
 {E_{{\bf q},{\bf p},k}^2}\right)
 \vphantom{-\frac{1}{2}\sum_{
 \begin{array}{c}
 \scriptstyle
 s_1,\ldots,s_5=\pm 1\\[-2.5mm]
 \scriptstyle
 s_1\ldots s_5=1
 \end{array}
 }\ln\left(1-\frac{2\omega_ls_1}{E_{{\bf q},{\bf p},k}}
 \right)}
 \right.\nonumber \\&&\left.
 -\,\frac{1}{2}\sum_{
 \begin{array}{c}
 \scriptstyle
 s_1,\ldots,s_5=\pm 1\\[-2.5mm]
 \scriptstyle
 s_1\ldots s_5=1
 \end{array}
 }\ln\left(1-\frac{2s_1\omega _1}
 {E_{{\bf q},{\bf p},k}}
 \right)
 \right].
 \end{eqnarray}
The last term may be rewritten as 
\be
\vphantom{\frac{1}{2}\sum_{
 \begin{array}{c}
 \scriptstyle
 s_1,\ldots,s_5=\pm 1\\[-2.5mm]
 \scriptstyle
 s_1\ldots s_5=1
 \end{array}
 }\ln\left(1-\frac{2\omega_ls_1}{E_{{\bf q},{\bf p},k}}
 \right)} \,\frac{1}{2}\sum_{
\begin{array}{c}
\scriptstyle  s_1,\ldots,s_5=\pm 1 \\ [-2.5mm] \scriptstyle  s_1\ldots s_5=1 
\end{array}
}\ln\left(1-\frac{2s_1\omega_l} {E_{{\bf q},{\bf p},k}}
\right) = 4 \ln\left(1-\frac{4\omega_l^2} {E^2_{{\bf q
},{\bf p},k}} \right). 
\ee

The sum over $k$ and the integrals over $q$ and $p$ then can be calculated
\begin{equation}\label{3.285} 
\int d^{l-1}q\,d^{l-1}p\, \sum_{k=0}^{\infty }\e^{-sE_{{\bf q},{\bf %
p},k}} =\left(\frac{\pi }{2s}\right)^{\frac{p}{2}-1}
\frac{\e^{-r^2s}}{\sinh(2\omega_ls)}. 
\end{equation}

We finally obtain the following form 
\begin{equation}\label{3.290} 
W= 
 \prod_{a=1}^{p-1}\left(
 \frac{n^{\frac{1}{l}}}{L_a}
 \right)
\int_0^{\infty} 
\frac{ds}{s}\left(\frac{\pi}{2s} \right )^{\frac{p}{2}-1} 
e^{-r^2s} 
\frac
{4\sinh^2(\omega_ls) - \sinh^2(2\omega_ls)} 
{~\sinh(2\omega_ls)}\,. 
\end{equation}

For large separation between branes we find: 
\begin{equation}\label{300}
W=
-4 (\frac{\pi}{2})^{\frac{p}{2}}  \,(3-l)!\,
\prod_{a=1}^{p-1}
 \left(
 \frac{n^{\frac{1}{l}}}{L_a}
 \right)
 \omega_l^3\frac 1{r^{8-p}}+ \cdot\cdot\cdot ,
\end{equation}
in agreement with supergravity results, especially for $p=2$ case \cite{DS,AB}.

\section{Conclusion}
In this article we studied some aspects of M-branes of \SMM.
First we introduced certain BPS configurations of the model, including
 single M$p$-brane ($p$=even) and two parallel M$p$-branes.
It was shown that these configurations do not take quantum corrections 
at one-loop.

Then the long range interactions of M$p$-branes were studied.
In Particular, we calculated the long range interaction between two
relatively rotated (in one and two angles) M$p$-branes and 
M$p$-anti-M$p$-branes. The results were in agreement with the 
11 dimensional supergravity. Moreover we 
found new BPS configurations in the two angle case which 
correspond to two M$p$-branes rotated with equal angles with respect
 to each other, as the same of \cite{SJ,TI}.

\noindent{\large\bf Acknowledgement}
\medskip

\noindent It is a pleasure to thank F. Ardalan for helpful discussions 
and reading the manuscript. We are also grateful
to M.M. Sheikh Jabbari and M. Khorrami for discussions. 


\end{document}